\documentclass[preprint,12pt,sort&compress]{elsarticle}

\usepackage{amssymb}
\usepackage{amsmath}

\usepackage{braket} 
\usepackage{algorithm,algorithmic} 
\usepackage{booktabs} 
\usepackage{array}    
\usepackage{caption}  

\journal{Expert Systems With Applications}

\begin{document}

\begin{frontmatter}



\title{An Amplitude-Encoding-Based Classical-Quantum Transfer Learning framework: Outperforming Classical Methods in Image Recognition} 


\author{Shouwei Hu, Xi Li, Banyao Ruan, Zhihao Liu} 

\affiliation{organization={Southeast University},
            addressline={}, 
            city={Nanjing},
            postcode={211189}, 
            state={Jiangsu},
            country={China}}

\begin{abstract}
The classical-quantum transfer learning (CQTL) method is introduced to address the challenge of training large-scale, high-resolution image data on a limited number of qubits (ranging from tens to hundreds) in the current Noisy Intermediate-Scale quantum (NISQ) era. existing CQTL frameworks have been demonstrate quantum advantages with a small number of parameters (around 50), but the performance of quantum neural networks is sensitive to the number of parameters. Currently, there is a lack of exploration into larger-scale quantum circuits with more parameters. This paper proposes an amplitude-encoding-based classical-quantum transfer learning(AE-CQTL) framework, accompanied by an effective learning algorithm. The AE-CQTL framework multiplies the parameters of quantum circuits by using multi-layer ansatz. Based on the AE-CQTL framework, we designed and implemented two CQTL neural network models: Transfer learning Quantum Neural Network (TLQNN) and Transfer Learning Quantum Convolutional Neural Network (TLQCNN). Both models significantly expand the parameter capacity of quantum circuits, elevating the parameter scale from a few dozen to over one hundred parameters. In cross-experiments with three benchmark datasets (MNIST, Fashion-MNIST and CIFAR10) and three source models (ResNet18, ResNet50 and DenseNet121), TLQNN and TLQCNN have exceeded the benchmark classical classifier in multiple performance metrics, including accuracy, convergence, stability, and generalization capability. Our work contributes to advancing the application of classical-quantum transfer learning on larger-scale quantum devices in future.
\end{abstract}



\begin{keyword}
Quantum machine learning \sep Parametrized quantum circuits \sep Quantum neural network \sep Quantum convolutional neural network
\sep Classical-quantum transfer learning


\end{keyword}

\end{frontmatter}

\section{Introduction}
Machine learning (ML), a cornerstone of artificial intelligence, focuses on enabling computers to uncover hidden features and relationships within data to perform specific tasks. Over the past two decades, rapid technological advancements have significantly enhanced our ability to collect, store, and share data. Leveraging its capacity to efficiently analyze and utilize vast amounts of data, ML has made remarkable strides, profoundly transforming our world. Its applications span from speech and image recognition to natural language processing (NLP) and recommendation systems \cite{lecun1989backpropagation,chan2016listen,vaswani2017attention}, permeating nearly every aspect of daily life. However, the explosive growth in data scale has introduced challenges for traditional classical ML algorithms. These include high computational resource consumption and prolonged training times when dealing with large-scale, high-dimensional datasets. For instance, the GPT-3 large language model \cite{brown2020language}, a milestone in NLP, comprises an astounding 175 billion parameters, requires 45TB of training data, and contributes to a training cost nearing one million dollars. To alleviate these limitations, quantum machine learning (QML) has arisen at the intersection of quantum computing and ML. QML leverages quantum phenomena, such as superposition and entanglement, to develop more efficient learning algorithms. For example, the quantum HHL algorithm \cite{harrow2009quantum} achieves exponential speedup over classical methods for solving linear equations. Similarly, Shor’s algorithm \cite{shor1999polynomial} demonstrates exponential acceleration for integer factorization, posing significant implications for cryptography, such as cracking RSA public-key encryption systems.\par

The variational quantum algorithm (VQA) \cite{cerezo2021variational}, based on parameterized quantum circuits (PQCs), has emerged as the main field in current quantum machine learning (QML) research. VQA is a quantum-classical hybrid algorithm in which PQCs is executed and measured on a quantum computer, and the classical expectation values obtained from the measurement are used to calculate the loss function and its gradient, enabling parameter optimization through a classical optimizer until convergence is achieved. Due to its structural and training similarities with classical neural networks, models based on VQA are often referred to as quantum neural networks (QNNs). This hybrid algorithm is particularly well-suited to the capabilities of current NISQ devices and holds promise for demonstrating quantum advantage \cite{preskill2018quantum}. VQA has garnered significant attention and has been applied to a wide range of problems, including quantum chemistry, quantum many-body physics, combinatorial optimization, and image classification. For instance, in 2014, Peruzzo et al. \cite{peruzzo2014variational} introduced the first application of VQA, the variational quantum eigensolver (VQE), designed for efficiently calculating the ground-state energy of molecules. Inspired by classical convolutional neural networks (CNNs), Cong et al. \cite{cong2019quantum} proposed the quantum convolutional neural network (QCNN) based on PQCs, which has been successfully applied to quantum many-body problems. Furthermore, Henderson et al. \cite{henderson2020quanvolutional} employed random quantum circuits as quantum convolutional layers to design a quantum-classical hybrid network for image recognition tasks, achieving accuracy surpassing that of classical CNNs. These advancements highlight the potential of VQA-based models to tackle complex problems and their promise for further expanding the applications of quantum computing in diverse domains.\par

It is particularly worth mentioning that an attractive research direction in the current VQA field is Classical-Quantum transfer learning (CQTL). This emerging quantum computing approach combines the strengths of classical deep learning models with quantum neural networks (QNNs), significantly reducing training costs and enhancing generalization capabilities. Mari et al. \cite{mari2020transfer} were the first to extend transfer learning, a widely adopted strategy in modern machine learning, to the hybrid classical-quantum domain, proposing the CQTL algorithmic framework. In essence, the CQTL framework leverages pre-trained classical neural networks as feature extractors, passing the extracted features to dressed quantum circuits (DQCs) \cite{mari2020transfer} for subsequent training and classification tasks. DQCs consist of a three-layer architecture: a pre-processing classical fully connected layer, a parameterized quantum circuit (PQC) layer, and a post-processing classical fully connected layer. Additionally, several studies based on this CQTL framework have achieved promising results in tasks such as image classification and speech recognition \cite{mogalapalli2022classical,qi2022classical,zhang2023remote,azevedo2022quantum}. However, a common issue in hybrid models based on the existing CQTL framework is the disproportionately small contribution of quantum parameters compared to classical parameters. Typically, PQCs in these models involve only around four qubits, making it challenging to distinguish the quantum component's impact from the classical component. This limitation runs counter to the primary goal of VQA research in the NISQ era: demonstrating quantum advantage. A similar discussion of the current CQTL framework was raised in a 2023 study by Kim et al. \cite{kim2023classical}. They proposed a convolutional neural network model utilizing CQTL and reported quantum advantages over classical models when the number of parameters was relatively small (around 50). However, their work provided limited analysis and validation of quantum advantages in scenarios involving larger parameter sets.\par

To address the limitations of the current CQTL framework, an amplitude-encoding-based CQTL (AE-CQTL) framework is introduced. In contrast to the original framework, this approach integrates a data embedding technique based on amplitude encoding, thereby eliminating the need for a preprocessing classical fully connected layer in dressed quantum circuits (DQCs). In this design, the quantum component assumes a primary role in high-level feature extraction, while the classical post-processing layer merely facilitates data dimensional transitions, emphasizing the pivotal role of parameterized quantum circuits (PQCs). Based on AE-CQTL framework, we propose a transfer learning quantum neural network (TLQNN) model tailored for image recognition tasks. The TLQNN model incorporates a universal repetitive ansatz structure in its PQCs, making it highly compatible with current NISQ devices. Furthermore, inspired by the QCNN architecture, we introduce a variant of the TLQNN model: the transfer learning quantum convolutional neural network (TLQCNN). The TLQCNN model features a quantum convolution layer and a quantum pooling layer within the PQCs, enabling secondary extraction and dimensionality reduction of data features encoded in the quantum Hilbert space. These features are subsequently processed through a quantum fully connected layer. This innovative design effectively mitigates the barren plateau phenomenon \cite{mcclean2018barren,cerezo2021cost} and demonstrates strong potential for scalability in future large-scale quantum devices.\par

The validation experiments were conducted using the Qiskit \cite{contributors2023qiskit} toolkit and the PyTorch \cite{paszke2019pytorch} neural network framework. Experimental results obtained from multiple datasets (including MNIST\cite{deng2012mnist}, Fashion-MNIST\cite{xiao2017fashion}, CIFAR10\cite{krizhevsky2009learning}) and various pre-trained classical networks (Res-Net\cite{he2016deep}, Dense-Net\cite{huang2017densely}) demonstrate the universality and suitability of the AE-CQTL framework. Moreover, to rigorously assess the performance of the AE-CQTL models, comparative experiments were conducted against benchmark classical transfer learning (CCTL) neural network models. The classical models consistently underperformed relative to the AE-CQTL models across all scenarios under identical training conditions, despite a higher number of training parameters. These results demonstrate the potential quantum advantage offered by the proposed AE-CQTL framework.\par

The remainder of this paper is organized as follows: Section 2 provides the main components of the quantum circuit in the CQTL framework. Section 3 details the proposed AE-CQTL framework and elaborates on the TLQNN and TLQCNN model structures. Section 4 describes the experimental design, datasets, training conditions, and provides an in-depth analysis of the results. Finally, Section 5 concludes the paper, summarizing the contributions and discussing future research directions and potential applications.\par

\section{Preliminaries}
\subsection{PQCs}
In general, PQCs is composed of multiple repeated layers of ansatz, each ansatz layer can be divided into training block and entanglement block, training block contains single qubit or multi-qubits parametric rotation gates \cite{benedetti2019parameterized}, entanglement block is composed of multi-qubits gates (such as CNOT gate, CZ gate) arranged according to a certain law. The single qubit gates and two-qubits gates \cite{nielsen2010quantum} used in this paper and their matrix representations are given in Table \ref{tab:quantum_gates}.  Let $\Theta=[\Theta_1,\Theta_2,\cdots,\Theta_l]$ represents the parameter vector of the $l$ layers in PQCs, and the training block and entanglement block of the $i$-th layer are represented by $U_i(\Theta_i)$ and nonparametric $E_i$ respectively, then PQCs has the form shown in Eq.\eqref{Eq:1}.\par
\begin{equation}\label{Eq:1}
U(\Theta) = \prod_{i=1}^{l}E_i U_i(\Theta_i)
\end{equation}

\begin{table}[h!]
  \centering
  \caption{The quantum gates used in this PQCs}
  \begin{tabular}{c c c}
      \toprule
      \textbf{Quantum gate} & \textbf{Symbol} & \textbf{Matrix representations} \\ 
      \midrule
      Hadamard & \( H \) & 
      \( \frac{1}{\sqrt{2}} 
      \begin{bmatrix}
          1 & 1 \\ 
          1 & -1
      \end{bmatrix} \) \\ 
      \addlinespace[1.5ex]
      \( R_x \) & \( R_x(\theta) \) & 
      \( \begin{bmatrix}
          \cos\frac{\theta}{2} & -i\sin\frac{\theta}{2} \\ 
          -i\sin\frac{\theta}{2} & \cos\frac{\theta}{2}
      \end{bmatrix} \) \\ 
      \addlinespace[1.5ex]
      \( R_y \) & \( R_y(\theta) \) & 
      \( \begin{bmatrix}
          \cos\frac{\theta}{2} & -\sin\frac{\theta}{2} \\ 
          \sin\frac{\theta}{2} & \cos\frac{\theta}{2}
      \end{bmatrix} \) \\  
      \addlinespace[1.5ex]
      \( R_z \) & \( R_z(\theta) \) & 
      \( \begin{bmatrix}
          e^{-i\frac{\theta}{2}} & 0 \\ 
          0 & e^{i\frac{\theta}{2}}
      \end{bmatrix} \) \\ 
      \addlinespace[1.5ex]
      \( U_3 \) & \( U_3 \) & 
      \( \begin{bmatrix}
          e^{i(\alpha-\frac{\beta}{2}-\frac{\delta}{2})} \cos\frac{\gamma}{2} & -e^{i(\alpha-\frac{\beta}{2}+\frac{\delta}{2})} \sin\frac{\gamma}{2} \\ 
          e^{i(\alpha+\frac{\beta}{2}-\frac{\delta}{2})} \sin\frac{\gamma}{2} & e^{i(\alpha+\frac{\beta}{2}+\frac{\delta}{2})} \cos\frac{\gamma}{2}
      \end{bmatrix} \) \\
      \addlinespace[1.5ex]
      Controlled NOT & CNOT & 
      \( \begin{bmatrix}
          1 & 0 & 0 & 0 \\ 
          0 & 1 & 0 & 0 \\ 
          0 & 0 & 0 & 1 \\
          0 & 0 & 1 & 0
      \end{bmatrix} \) \\
      \bottomrule
  \end{tabular}
  \label{tab:quantum_gates}
\end{table}

\begin{figure}[t]
  \centering
  \includegraphics[scale=1.3]{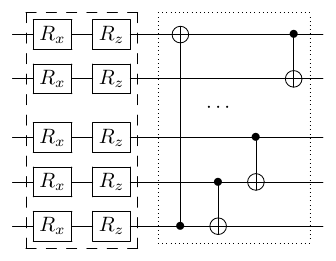}
  \caption{An ansatz used in PQCs. (The dashed box is the training block, and the dotted box is the entanglement block.)}
  \label{fig:1}
\end{figure}

Due to their multi-layer structure resembling classical feedforward neural networks \cite{goodfellow2016deep}, quantum models based on PQCs are often referred to as quantum neural networks (QNNs). These models have demonstrated excellent performance across various applications \cite{perdomo2018opportunities,peruzzo2014variational,mcclean2018barren}. Theoretically, as the number of ansatz layers in PQCs increases, their expressive capacity also grows, which is known as the Universal Approximation Property of PQCs \cite{goto2021universal,perez2021one,yu2022power}. This property is analogous to the Universal Approximation Theorem in classical machine learning theory \cite{cybenko1989approximation,hornik1991approximation}. For instance, Fig.\ref{fig:1} illustrates the structure of an ansatz employed in the PQCs \cite{sim2019expressibility}, the dashed box represents the training block, while the dotted box indicates the entanglement block.\par

\subsection{Data embedding and measurement}
Data embedding refers to the process of mapping classical data from a classical vector space to quantum Hilbert space, which is simply to encode a real vector $\mathbf{x}$ in the classical domain into a quantum state $\ket{f(\mathbf{x})} $. This step is fundamental in quantum machine learning (QML) tasks and is an indispensable component for most quantum algorithms based on parameterized quantum circuits (PQCs). The choice of a suitable data embedding strategy is crucial, as it directly influences the generalization performance of the algorithm \cite{schuld2020circuit,cong2019quantum,grant2018hierarchical,ostaszewski2021reinforcement}. Existing studies have demonstrated the impact of embedding strategies from multiple perspectives, including quantum information theory and kernel methods \cite{caro2021encoding,banchi2021generalization,huang2021power,liu2021rigorous,schuld2021quantum,perez2020data,schuld2021effect,li2022concentration}. Currently, two of the most widely adopted data embedding strategies are angle encoding and amplitude encoding.\par
\begin{figure}[t]
  \centering
  \includegraphics[scale=0.2]{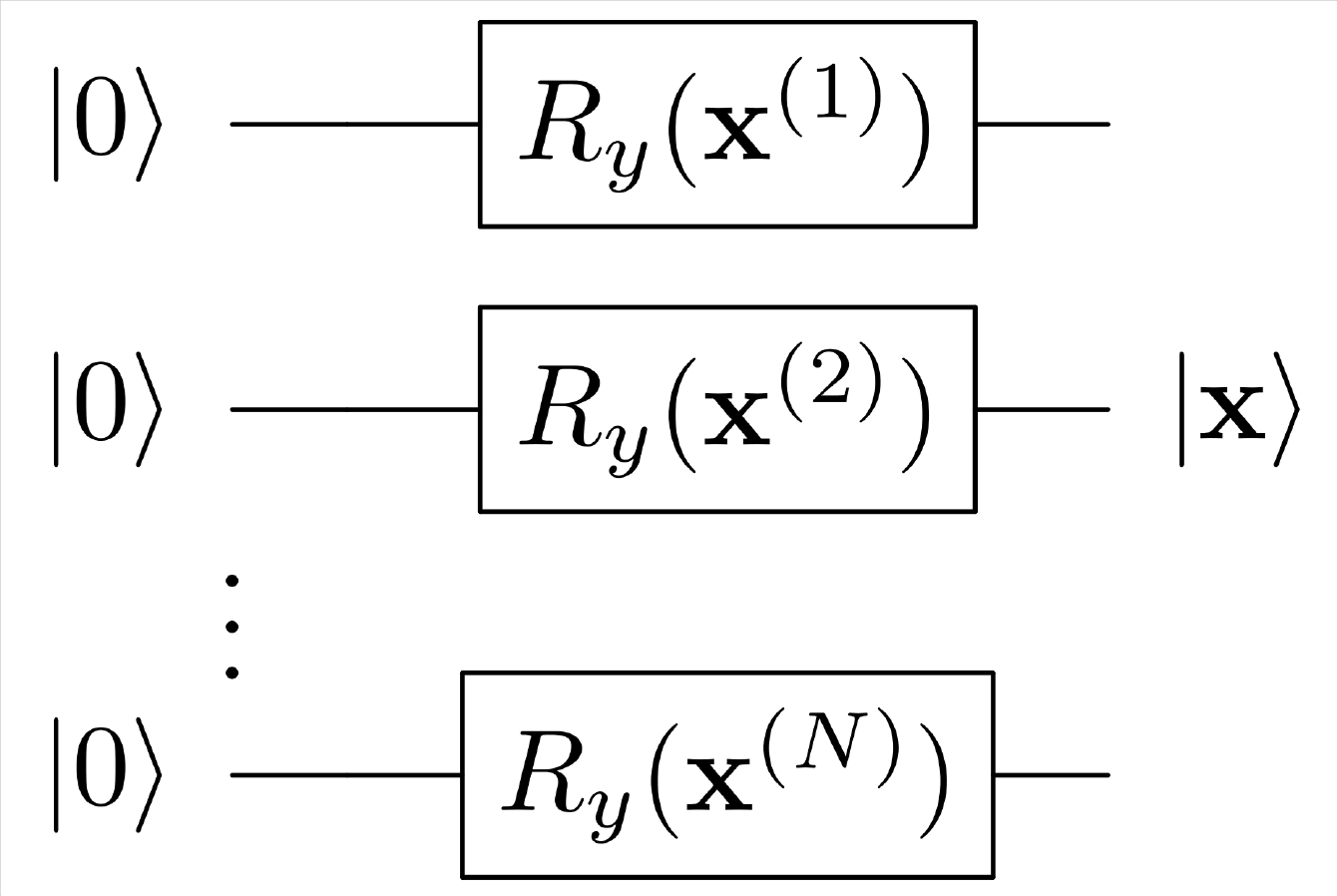}
  \caption{Schematic diagram of angle coding in $N$ qubits}
  \label{fig:2}
\end{figure}

The schematic diagram of angle encoding is shown in Fig.\ref{fig:2}. Let the classical real number vector is $ \mathbf{x}=[\mathbf{x}^{(1)},\cdots, \mathbf{x}^{(N)}] \in R^N $, then $N$ qubits are required for encoding it under the angle encoding strategy. For example, by applying $R_y(\mathbf{x}^{(i)})$ to the $i$-th $(1\le i \le n)$ qubit with an initial state of $\ket{0}$, the quantum state $\ket{\mathbf{x}}$ shown in Eq.\eqref{Eq:2} can be obtained.\par
\begin{equation}\label{Eq:2}
  \ket{\mathbf{x}} = \otimes_{i=1}^{N} cos \left( \mathbf{x}^{(i)} \right) \ket{0} + sin \left( \mathbf{x}^{(i)} \right) \ket{1}
\end{equation}

The angle encoding strategy often requires preprocessing classical data, such as employing interpolation, principal component analysis (PCA), or autoencoder algorithms to reduce the dimensionality of the data. However, these preprocessing steps inevitably introduce a degree of information loss when transforming the original input data. In contrast, amplitude encoding offers a different approach and is widely utilized in areas like quantum image representation and quantum machine learning (QML). This strategy associates the normalized input data with the amplitudes of a quantum state. Compared to angle encoding, amplitude encoding has a significant advantage: it requires only a logarithmic number of qubits relative to the length of the classical data vector. This efficiency reduces the quantum resource requirements, making it especially suitable for high-dimensional data. The quantum state encoded using amplitude encoding can be expressed as shown in Eq.\eqref{Eq:3}, where $\ket{i}$ represents the $i$-th computational basis state.\par
\begin{equation}\label{Eq:3}
  \ket{\mathbf{x}} = \frac{1}{\Vert \mathbf{x} \Vert} \sum_{i=0}^{N} \mathbf{x}^{(i+1)} \ket{i}
\end{equation}
Different from data embedding, measurement refers to the process of mapping the final state of a quantum circuit from quantum Hilbert space back into the classical vector space. Similar to nonlinear activation functions (such as ReLU, Tanh, and Sigmoid) are introduced in classical neural networks, the measurement of quantum circuits will reduce the degree of freedom of the entire quantum system, thus introducing nonlinearity. Research have shown this approach can enhance the expressive capacity of QNNs based on parameterized quantum circuits (PQCs)\cite{cong2019quantum,henderson2020quanvolutional,kim2023classical,huang2017densely}.\par
Pauli-Z measurement is to take Pauli-Z operator as observable, that is, to take $\{\ket{0}\bra{0}_1,\ket{1}\bra{1}_{-1} \}$ as measurement operator, and to obtain the expectation value under the index $\{ 1,-1 \}$.Taking a single qubit system with state $\ket{\varPsi}=a\ket{0}+b\ket{1}$ as an example, where $a$ and $b$ are real numbers, then the result measured by Pauli-Z can be expressed as $\bra{\varPsi}Z\ket{\varPsi} = a^2-b^2$. Generally, $\bra{\varPsi}Z\ket{\varPsi}$ can be abbreviated as $\braket{Z}$.\par
For a quantum system with $n$ qubits, the observable used to measure the $k$-th qubit $(1 \le k \le n)$ can be expressed as $Z_k = I^{\otimes k-1}\otimes Z \otimes I^{\otimes n-k}$. Let $\ket{\mathbf{x}}$ be a quantum state encoded by classical data. PQCs are represented by $U(\Theta)$ and apply to $\ket{\mathbf{x}}$, and the measurement result of the final state $\ket{\varPsi(\Theta)} = U(\Theta)\ket{\mathbf{x}}$ can be expressed as a classical data vector $\mathbf{y}=\left[ \mathbf{y}^{(1)},\cdots,\mathbf{y}^{(n)} \right] \in R^{n}$, where the element $\mathbf{y}^{(k)}(1 \le k \le n)$ has the form shown by Eq.\eqref{Eq:4}.\par
\begin{equation}\label{Eq:4}
  \ket{\mathbf{y}^{(k)}} = \bra{\mathbf{x}} U^{\dagger}(\Theta)Z_k U(\Theta)\ket{\mathbf{x}} = \bra{\mathbf{x}} U^{\dagger}(\Theta)I^{\otimes k-1}\otimes Z \otimes I^{\otimes n-k}  U(\Theta)\ket{\mathbf{x}}
\end{equation}
In addition, the purpose of adopting local observable instead of global observable is to alleviate the impact of barren plateau \cite{mcclean2018barren,cerezo2021cost}, reduce the training difficulty and improve the trainability.\par

\subsection{Transfer learning}
Transfer learning has achieved successful applications in various fields such as computer vision, NLP, speech recognition, and generative models \cite{pan2009survey,tan2018survey,yosinski2014transferable}. Its primary advantage lies in its ability to significantly accelerate the training process of new models, reduce the consumption of computing resources and time, and achieve excellent performance even with small sample datasets by leveraging pre-trained models (referred to as source models).\par

Depending on the types of source models and target models (classical or quantum), transfer learning can be classified into four types: CCTL (classical-to-classical transfer learning), CQTL (classical-to-quantum transfer learning), QCTL (quantum-to-classical transfer learning), and QQTL (quantum-to-quantum transfer learning). For instance, CQTL stands for classical-to-quantum transfer learning, which involves utilizing classical source model as feature extractor and replacing the final few layers (fully-connected layer) of the source model with PQCs to be trained.\par

In the current NISQ era, CQTL holds higher practical value compared to QCTL and QQTL. This is because CQTL effectively combines the advantages of state-of-the-art, practically proven deep neural networks (such as ResNet \cite{he2016deep}, Inception \cite{szegedy2015going}, DenseNet \cite{huang2017densely}) with quantum circuits, offering faster training speeds and enhanced noise resilience. This advantage has been verified on currently available quantum computers, such as ibmqx4 and Aspen-4-4Q-A \cite{mari2020transfer}.\par

\section{The Amplitude-Encoding-Based CQTL framework and the corresponding models}
The concept of CQTL was originally introduced by Mari et al. in 2020 \cite{mari2020transfer}. Its core idea involves utilizing pre-trained classical models as feature extractors and incorporating DQCs (dressed quantum circuits) to construct a quantum-classical hybrid network. Taking the CQTL application example provided by Mari et al., the DQCs is designed as $L_{512\rightarrow 4}\rightarrow Q \rightarrow L_{4\rightarrow 2}$, where $Q$ represents a PQCs with 4 qubits and 6 layers of $ansatz$, and $L$ denotes the classical fully-connected network. However, this structure is not enough to demonstrate the quantum advantage of CQTL mode. Firstly, the number of classical parameters far exceeds that of the quantum rotation gate parameters. In DQCs, quantum parameters constitute less than one percent of the total classical parameters. Secondly, since the classical network $L_{4\rightarrow 2}$ performs dimension reduction and feature extraction on the data before passing it to the PQCs, it becomes difficult to clearly isolate and evaluate the contribution of PQCs in accomplishing the task. These limitations have also been observed in other CQTL studies based on DQCs \cite{mogalapalli2022classical,qi2022classical,zhang2023remote,azevedo2022quantum}.\par
To harness the advantages of quantum computing as much as possible, we propose a amplitude-encoding-bases CQTL framework and its corresponding learning algorithm in this section. This framework substantially reduces the reliance on classical parameters, enabling a clearer demonstration of quantum advantages compared to existing approaches. Building on this foundation, we introduce a CQTL neural network model designed specifically for NISQ devices, termed TLQNN. Additionally, we present a variant of TLQNN, named TLQCNN, which integrates transfer learning techniques with quantum convolutional concepts. The design demonstrates its potential and value in implementing quantum machine learning applications on large-scale quantum devices.\par

\subsection{The amplitude-encoding-based CQTL framework}
\begin{figure}[t]
  \centering
  \includegraphics[scale=0.29]{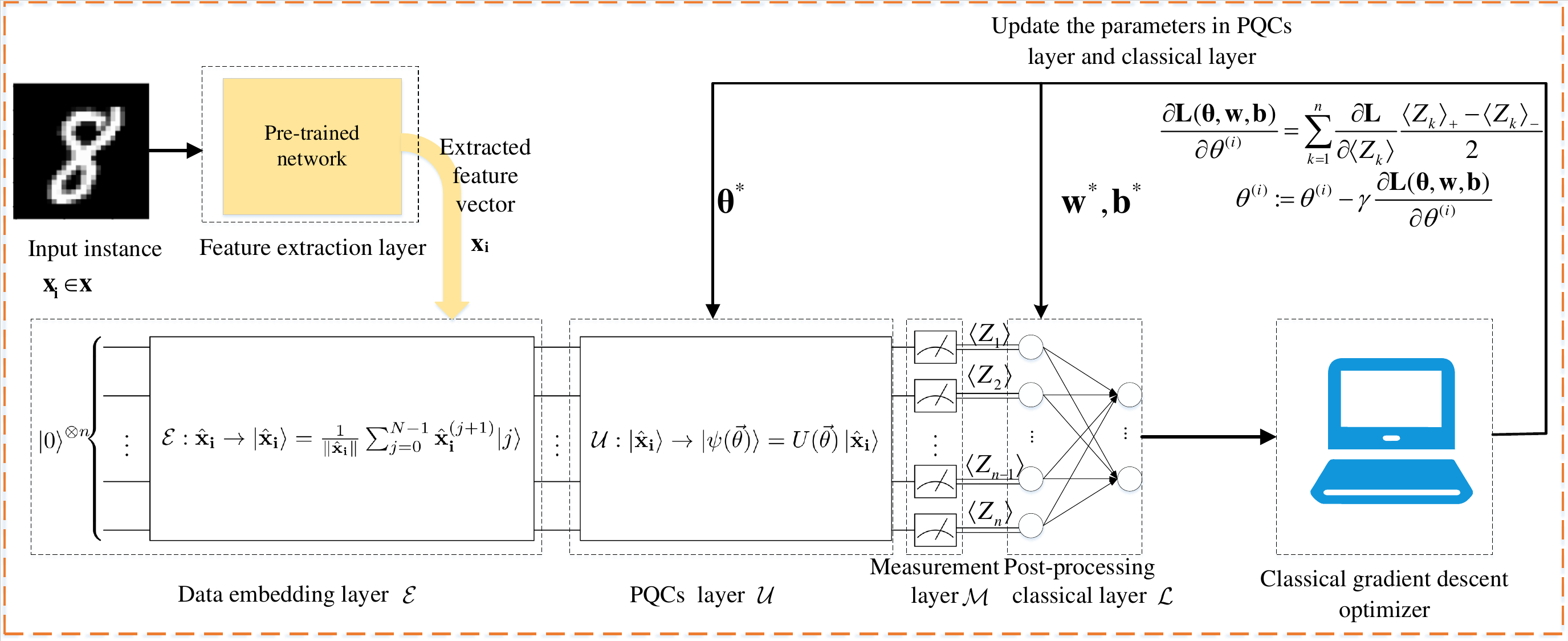}
  \caption{The proposed AE-CQTL framework structure diagram based on amplitude encoding. It is mainly composed of five parts: feature extraction layer, data embedding layer, PQCs layer, quantum measurement layer, and post-processing classical network layer.}
  \label{fig:3}
\end{figure}

This subsection provides a detailed description of the proposed AE-CQTL framework and its learning algorithm. As depicted in Fig.\ref{fig:3}, the framework comprises five core layers, arranged in the order of data instance flow: feature extraction layer $\mathcal{A}$, data embedding layer $\mathcal{E}$, PQCs layer $\mathcal{U}$, measurement layer $\mathcal{M}$, and post-processing classical network layer $\mathcal{L}$. Assuming the input data instances are images, they first undergo pre-processing steps such as resizing and normalization. The pre-processed images then fed into the feature extraction layer, which utilizes pre-trained models to extract initial features, converting the images into feature vectors and effectively reducing data dimensionality. Subsequently, these classical feature vectors are passed to the data embedding layer, where they are encoded as quantum states in the quantum Hilbert space using amplitude encoding techniques. Following this, the PQCs layer, equipped with trainable parameters, acts on the amplitude-encoded quantum states to perform advanced feature extraction. The resulting quantum final state encodes distinctive features that differentiate the input instance from others. The measurement layer then converts the feature information encoded in the quantum state into classical information through measurement operations under given observables. Finally, the measurement results are fed into the post-processing classical network layer, whose output is used to calculate the loss value and optimize the trainable parameters within the framework using a classical gradient descent optimizer, ultimately leading to a classification result. It's worth emphasizing that the post-processing classical network layer introduces only a small number of classical parameters, which essentially perform simple linear processing on the PQCs measurement results, without any nonlinear activation functions. Its main role is to enable the framework's flexible application to multiclass classification problems without overshadowing the core role of the quantum components.\par
Here, we delve into the learning algorithm of the proposed AE-CQTL framework, as outlined below. It is worth emphasizing that during the execution of algorithm, parameter updates are solely conducted for a single training sample in each iteration. Nevertheless, this single-sample-based updating strategy can be seamlessly extended to incorporate a cumulative loss minimization approach, accommodating various optimization requirements.\par

\begin{algorithm}[!ht]
  \renewcommand{\algorithmicrequire}{\textbf{Input:}}
	\renewcommand{\algorithmicensure}{\textbf{Output:}}
	\caption{Learning Algorithm of The Proposed AE-CQTL Framework}
    \begin{algorithmic}[1] 
      \REQUIRE Training set: $T={(\mathbf{x_1},\mathbf{y_1}),(\mathbf{x_2},\mathbf{y_2}),\cdots,(\mathbf{x_M},\mathbf{y_M})}$, learning rate: $\gamma$, qauntum parameter vector: $\Theta= [ \theta^{(1)},\cdots,\theta^{(m)} ]$, classical weights: $\mathbf{w}$, classical bias: $\mathbf{b}$, maximum epochs: $\varepsilon$ ;
	    \ENSURE Optimized parameters: $\Theta^{\star}, \mathbf{w}^{\star}, \mathbf{b}^{\star}$; 
      
        \STATE Randomly initialize all parameters including: $\Theta, \mathbf{w}, \mathbf{b}$
        \REPEAT 
          \FORALL {$(\mathbf{x}_i,\mathbf{y}_i)\in T$}

            \STATE Input $\mathbf{x}_i$ into the source model $\mathcal{A}$ to obtain the corresponding feature vector $\hat{\mathbf{x}}_i \in R^{n_{in}}$.

            \STATE The feature vector $\hat{\mathbf{x}}_i$ is filled to lenth $N=2^{\lceil log_2 n_{in} \rceil}$, and then $\ket{\hat{\mathbf{x}}_i}=\frac{1}{\Vert \hat{\mathbf{x}}_i \Vert} \sum_{j=0}^{N-1}\hat{\mathbf{x}}_i^{(j+1)} \ket{j}$ is obtained after amplitude encoding through the $n$-qubit $(n=log_2 N)$ quantum circuit of the data embedding layer $\mathcal{E}$.

            \STATE PQCs $U(\Theta)$ act on the quantum state $\ket{\hat{\mathbf{x}}_i}$, to obtain final states $\ket{\varPsi(\Theta)}=U(\Theta)\ket{\hat{\mathbf{x}}_i}$.

            \STATE The final state $\ket{\varPsi(\Theta)}$ is measured. The number of qubits to be measured can be adjusted in different cases, and it is assumed that all qubits are measured separately by Pauli-Z, resulting in a measurement result $[ \braket{Z_1},\cdots,\braket{Z_n} ]$ of length $n$ consisting of expectation values.

            \STATE Input the measurement result into the classical fully-connected layer $\mathcal{L}$ and calculate the
            loss value $\mathbf{L(\Theta, \mathbf{w}, \mathbf{b})}$ based on the outputs of $\mathcal{L}$ and data label $\mathcal{y}_i$.

            \STATE Update parameters and the update rules as shown in Eq. (5). Where $\partial_{\mathbf{w}}(\Theta, \mathbf{w}, \mathbf{b})$ and $\partial_{\mathbf{b}}(\Theta, \mathbf{w}, \mathbf{b})$ are given by automatic difference frameworks such as PyTorch. $\partial_{\Theta^{(i)}}(\Theta, \mathbf{w}, \mathbf{b})$ is generally calculated using the chain rule combined with the parameter-shift rule \cite{schuld2019evaluating,crooks2019gradients} (Eq.\eqref{Eq:5}), and can also be estimated by the finite difference method.
          \ENDFOR
          \UNTIL { the maximum epochs $\varepsilon$ is reached }
        \STATE \textbf{return} $\Theta^{\star}, \mathbf{w}^{\star}, \mathbf{b}^{\star}$.
    \end{algorithmic}
\end{algorithm}

\clearpage

\begin{equation}\label{Eq:5}
\begin{aligned}
\theta^{(i)} &= \theta^{(i)}-\gamma \frac{\partial \mathbf{L}(\Theta, \mathbf{w}, \mathbf{b})}{\partial \theta^{(i)}}\\
\mathbf{w} &= \mathbf{w}-\gamma \frac{\partial \mathbf{L}({\Theta}, \mathbf{w}, \mathbf{b})}{\partial \mathbf{w}}\\
\mathbf{b} &= \mathbf{b}-\gamma \frac{\partial \mathbf{L}({\Theta}, \mathbf{w}, \mathbf{b})}{\partial \mathbf{b}}
\end{aligned}
\end{equation}

\begin{equation}\label{Eq:6}
\frac{\partial \mathbf{L}(\Theta, \mathbf{w}, \mathbf{b})}{\partial \theta^{(i)}}=\sum_{k=1}^{n} \frac{\partial \mathbf{L}}{\partial\braket{Z_{k}}} \frac{\partial\braket{Z_{k}}}{\partial \theta^{(i)}}=\sum_{k=1}^{n} \frac{\partial \mathbf{L}}{\partial \braket{Z_{k}}} \frac{\braket{Z_{k}}_{+}-\braket{Z_{k}}_{-}}{2}
\end{equation}
In Eq.\eqref{Eq:6}, $\braket{Z_{k}}_{\pm}$ represents the expectation value measured by shifting $\theta^{(i)}$ by $\pm \frac{\pi}{2}$.

\subsection{The TLQNN model}
\begin{figure}[t]
  \centering
  \includegraphics[scale=0.17]{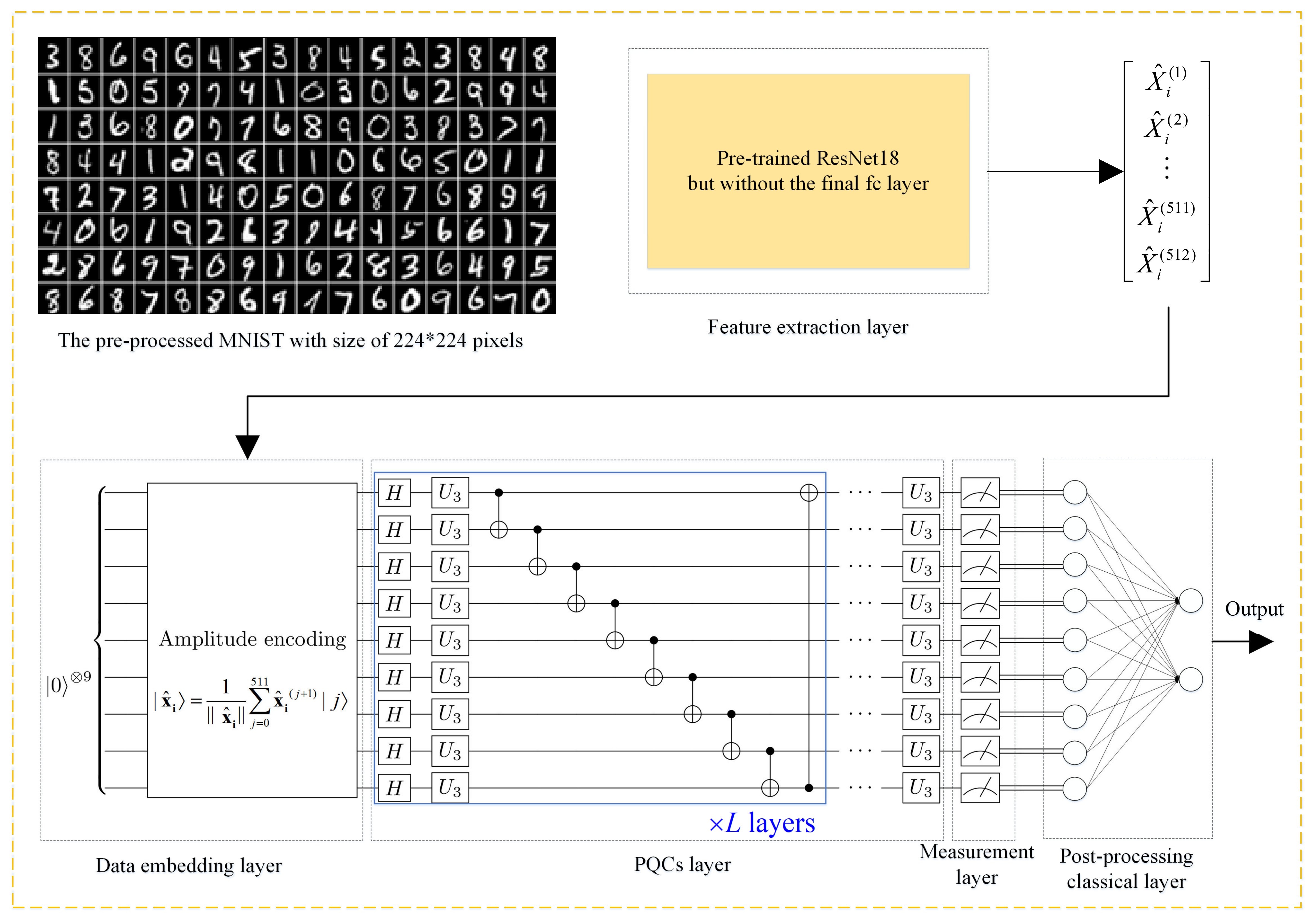}
  \caption{The architecture of TLQNN with ResNet18 as the source model.}
  \label{fig:4}
\end{figure}

Based on the AE-CQTL framework proposed in Section 3.1, we have designed the transfer learning quantum neural network (TLQNN) model for image classification. Taking the binary classification between digits '3' and '8' from the MNIST dataset as an example, we present the specific architecture of TLQNN with ResNet18 as the source model in Fig.\ref{fig:4}. The black dashed boxes in the figure correspond to the five components introduced in the AE-CQTL framework.\par
As depicted in Fig.\ref{fig:4}, during the pre-processing stage, to meet the input size requirements of the pre-trained model, we resize the original $28\times 28$ pixel MNIST handwritten images to $224\times 224$ pixels and feed them into the pre-trained ResNet18 model. This source model has been pre-trained on the ImageNet, with the final fully-connected layer $\mathcal{L}_{512\rightarrow 100}$ removed. Using the ResNet18 source model, we extract a 512-dimensional feature vector, which is then encoded into a 9-qubits quantum circuit using amplitude encoding. The PQCs layer comprises repeated $L$ times of 9-qubits ansatz layer (see the blue solid-lined box). The training block consists of a sequence of $H$ gates and $U_3$ gates, with the inclusion of $U_3$ gates enhancing the generality of the training block's design. The entanglement block uses CNOT gates to establish loop entanglement. This circular entanglement structure, also referred to as a cyclic code entanglement block in \cite{schuld2020circuit}, maximizes bipartite entanglement entropy. Systematic studies have highlighted that the topology of PQCs significantly influences both the expressive capacity and entanglement potential of the model \cite{sim2019expressibility}. Notably, before the measurement layer, we introduce an additional layer of $U_3$ gates, an empirical approach aimed at enhancing model performance. The measurement layer performs Pauli-Z measurements on all qubits, yielding classical expectation values $\{\braket{Z_k}\}_{k=1}^{9}$. On one hand, we hope to leverage as much information as possible from the quantum circuit through this strategy. On one hand, we aim to extract as much information as possible from the quantum circuit. On the other hand, Pauli-Z measurements, based on local rather than global observables, help mitigate the barren plateau phenomenon. The resulting measurement outcomes are then passed through a post-processing classical fully-connected layer $\mathcal{L}_{9\rightarrow 2}$ to produce the final classification results.\par
The TLQNN model, based on the AE-CQTL framework, exhibits unique flexibility in terms of parameter quantity. Specifically, it boasts a total of $27(L+1)$ quantum parameters and only 20 classical parameters. When the number of ansatz layers is sufficiently large, the quantum parameters surpass the classical ones, highlighting the quantum advantage of the model. The structure of this model clearly demonstrates that the post-processing classical fully-connected layer plays a supporting role, emphasizing the centrality of the quantum component. Additionally, the design of the TLQNN model offers remarkable flexibility, enabling adjustments based on the chosen source model. For instance, if a pre-trained DenseNet121 is employed as the source model, given its feature length of 1024, a 10-qubit circuit will be utilized for the data embedding layer and PQCs layer, increasing the quantum parameter count to $30(L+1)$. Alternatively, if a pre-trained ResNet50 is selected, with a feature length of 2048, a 11-qubit data embedding layer and PQCs layer will be employed, resulting in a quantum parameter count of $33(L+1)$. Throughout this process, the number of input nodes in the post-processing classical fully-connected layer is also adjusted to match the changing number of qubits. This flexibility allows the TLQNN model to adapt seamlessly to different pre-trained models and achieve efficient transfer learning in various scenarios.\par

\subsection{A special variant of the TLQNN: TLQCNN}
The PQCs layer of the TLQNN model employs a repeatedly stacked ansatz structure, which demonstrates excellent adaptability and universality on current NISQ devices. Typically, the performance of the model can be further enhanced by increasing the number of ansatz layers. This simple and efficient circuit construction paradigm is highly appealing in the NISQ era and has therefore been adopted by most variational quantum models. However, it is crucial to acknowledge the limitations inherent in the PQCs construction paradigm. On one hand, the existence of expressibility saturation \cite{sim2019expressibility} restricts the effective number of layers in PQCs, limiting the continuous improvement of model performance solely through layer addition. On the other hand, as future large-scale quantum devices emerge, with the increase in available qubits and circuit depth, variational quantum models are more likely to encounter the barren plateau phenomenon, characterized by an extremely flat cost function landscape. This will render gradient-based learning algorithms ineffective, making model training challenging. Despite various efforts to address this issue through heuristic parameter initialization strategies \cite{grant2019initialization} and learning methods \cite{skolik2021layerwise}, a definitive solution to the barren plateau phenomenon remains elusive.\par
Recently, a study on quantum convolutional neural network (QCNN) has revealed a promising insight. Unlike the prevalent PQCs architectures based on repeated ansatz layers, QCNN demonstrate a remarkable resilience against the barren plateau phenomenon. Even with random parameter initialization, the gradient descent speed of QCNN does not exceed a polynomial level with respect to the number of qubits \cite{pesah2021absence}. This significant advantage ensures the trainability of QCNN, overcoming a challenging issue faced by many variational quantum models. Furthermore, existing research has shown that QCNN, despite having fewer parameters, often exhibit superior performance and generalization capabilities compared to classical algorithms \cite{hur2022quantum}. Motivated by these findings, and building upon the proposed AE-CQTL framework introduced in Section 3.1, here we present a groundbreaking integration of the QCNN with the proposed framework, giving birth to a novel quantum-classical hybrid convolutional neural network model for transfer learning: TLQCNN.\par
\begin{figure}[t]
  \centering
  \includegraphics[scale=0.2]{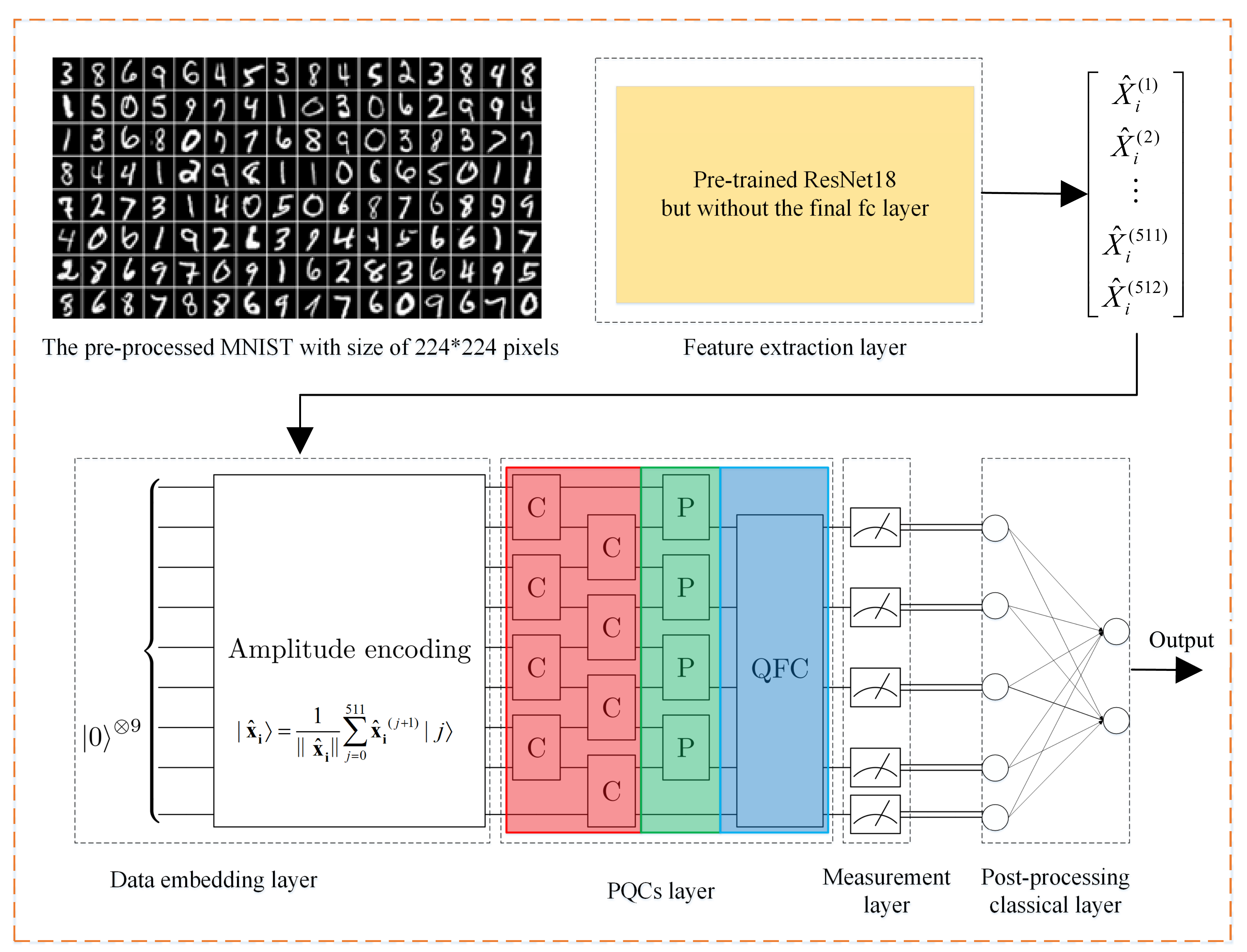}
  \caption{The schematic diagram of the TLQCNN structure with ResNet18 as the source model. The red shaded part of the PQCs layer is the quantum convolution layer, the green shaded part is the quantum pooling layer, and the blue part is the quantum fully connected layer.}
  \label{fig:5}
\end{figure}

The proposed TLQCNN model adopts a fundamentally different strategy for its PQCs layer. Specifically, the TLQCNN model leverages a quantum convolutional layer to perform secondary extraction of data features encoded in the quantum Hilbert space. The core idea behind this approach is that quantum convolutional layer can more efficiently utilize kernel methods in high-dimensional Hilbert space compared to classical computers, as indicated by \cite{schuld2021quantum}. Subsequently, a quantum pooling layer discards unnecessary information and retains more critical features on a smaller number of qubits. Based on this, TLQCNN only applies a quantum fully-connected layer to the quantum subsystem composed of these few qubits, followed by a measurement operation to map the feature information in the quantum Hilbert space to classical information. This PQCs structure not only guarantees the performance of the TLQCNN model but also effectively mitigates the impact of the barren plateau phenomenon, thereby ensuring the trainability and practical application potential of the model. Using the binary classification of digits '3' and '8' from the MNIST dataset as an example, Fig.\ref{fig:5} illustrates the schematic diagram of the TLQCNN structure with ResNet18 as the source model.\par
\begin{figure}[t]
  \centering
  \includegraphics[scale=0.23]{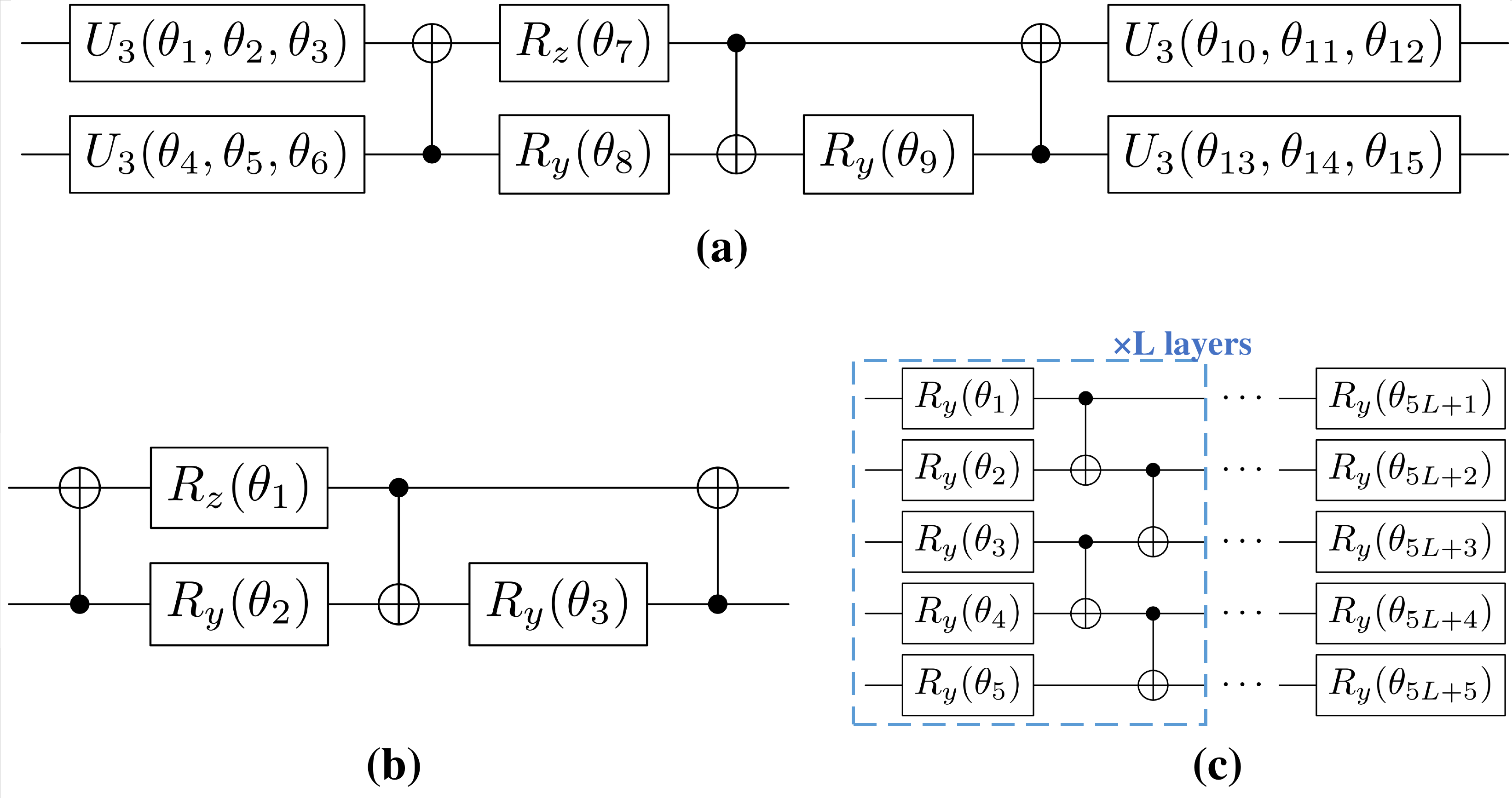}
  \caption{The schematic diagram of the TLQCNN structure with ResNet18 as the source model. The red shaded part of the PQCs layer is the quantum convolution layer, the green shaded part is the quantum pooling layer, and the blue part is the quantum fully connected layer.}
  \label{fig:6}
\end{figure}

In Fig.\ref{fig:5}, the five components of the AE-CQTL framework are outlined by the black dashed boxes. The feature extraction layer and data embedding layer remain consistent with the previous design, thus they will not be further discussed here. The most critical adjustment in TLQCNN compared to TLQNN lies in the PQCs layer. Within the PQCs layer, we introduce a quantum convolutional and a quantum pooling layer, corresponding to the red and green shaded area in Fig.\ref{fig:5}, respectively. The quantum convolutional layer is specifically implemented by consecutively applying the convolutional operator $\textbf{C}$, as shown in Fig.\ref{fig:6}(a), to all adjacent qubits. The quantum pooling layer employs the pooling operator $\textbf{P}$ depicted in Fig.\ref{fig:6}(b), where the qubit located on the top is discarded by tracing out, and after effective training of the model, critical feature information is retained on the qubit below. For instance, in the 9-qubits TLQCNN model depicted in Fig.\ref{fig:5}, only qubits $\{q_1,q_3,q_5,q_7,q_8\}$ are retained after the quantum pooling layer, and subsequent operations are performed solely on the subsystem composed of these rest of qubits.\par
\begin{figure}[t]
  \centering
  \includegraphics[scale=0.4]{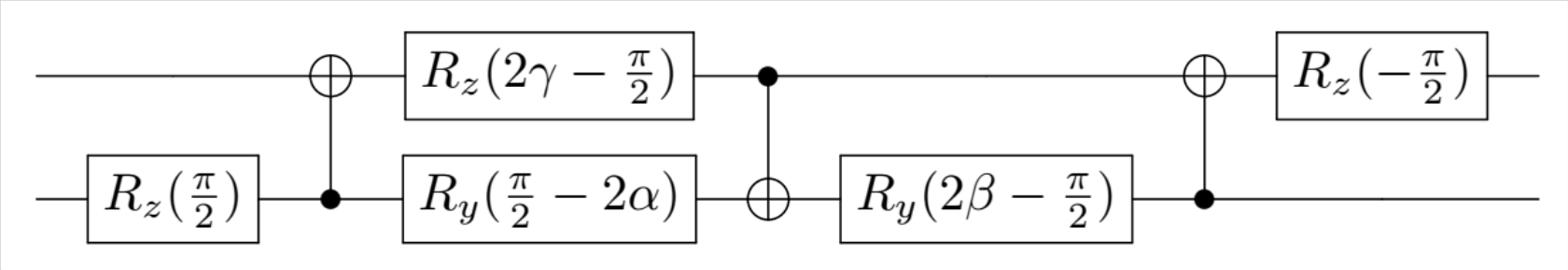}
  \caption{The quantum circuit of $N(\alpha,\beta,\gamma)$.}
  \label{fig:7}
\end{figure}

Indeed, the choice of convolutional and pooling operators is diverse, as already discussed in \cite{lecun1989backpropagation}. It's worth noting that the circuit construction of the convolutional operator shown in Fig.\ref{fig:6}(a) is designed with considerations for both generality and universality. Specifically, this circuit represents the optimal structure for all two-qubits unitary gates (more rigorously, $SU(4)$ unitary gates) \cite{vatan2004optimal}, requiring a maximum of 3 CNOT gates and 15 Pauli rotation gates. A more formal expression of the convolutional operator is given by Eq.\eqref{Eq:7}.
\begin{equation}\label{Eq:7}
C=(A_1 \otimes A_2)\cdot N(\alpha,\beta,\gamma)\cdot (A_3\otimes A_4)
\end{equation}
where $A_j\in SU(2)$, that is $U_3$ gate, $N(\alpha,\beta,\gamma)$ is shown in Eq.\eqref{Eq:8}.
\begin{equation}\label{Eq:8}
N(\alpha,\beta,\gamma) = exp \big[i(\alpha \sigma_x \otimes \sigma_x + \beta \sigma_y \otimes \sigma_y + \gamma \sigma_z \otimes \sigma_z) \big]
\end{equation}
At the same time, Fig.\ref{fig:7} shows the specific quantum circuit diagram of $N(\alpha,\beta,\gamma)$.
The pooling operator $\textbf{P}$ used in this paper is actually a simplified implementation of the $N(\alpha,\beta,\gamma)$ circuit, where the parameterized Pauli rotation gates and three two-qubit CNOT gates are preserved, while the Pauli-Z rotation gates on both ends are removed. More specifically, $R_z(-\pi/2)$ is discarded along with the upper qubit since it does not have any effect on the lower qubit. On the other hand, $R_z(-\pi/2)$ is absorbed by the $U_3$ gate acting on the same qubit in the preceding quantum convolutional layer. The design philosophy behind the pooling operator $\textbf{P}$ in this paper is to retain as much expressivity as possible from the convolutional operator $\textbf{C}$ while having fewer parameters and a simpler circuit structure.\par
After the quantum pooling layer, TLQCNN applies a quantum fully-connected layer to process the feature information obtained from the quantum pooling operation. Since the feature extraction has already been handled by the quantum convolutional layer, the quantum fully-connected layer does not require the complex ansatz structure used in the TLQNN model. Instead, a simpler structure is adopted in this paper, as shown in Fig.\ref{fig:6} (c). Specifically, the training block consists of a sequence of Pauli-Y rotation gates, and the entanglement block uses CNOT gates to create linear entanglement. Additionally, this paper empirically adds an extra layer of Pauli-Y rotation gates before the measurement layer to potentially enhance model performance. In the measurement layer, only the subsystem composed of qubits $\{q_1,q_3,q_5,q_7.q_8\}$ is measured using Pauli-Z operators, and the expectation values are passed to the post-processing classical fully-connected layer $\mathcal{L}_{5\rightarrow 2}$ for linear processing, ultimately yielding the final classification result. From the perspective of the model structure, the quantum convolutional layer and quantum pooling layer are the key contributors to the performance of the TLQCNN model, while the post-processing classical fully connected layer merely serves as a bridge for connecting dimensions and performing linear processing of quantum measurement results.\par
When equipped with only one layer of quantum convolution and pooling, the number of parameters in each part of $n$-qubit TLQCNN model is distributed as follows. Specifically, the quantum convolutional layer comprises $15(n-1)$ parameters, the quantum pooling layer holds $3\times \lfloor n/2 \rfloor$ parameters, and the quantum fully- connected layer has $(L+1)\times \lceil n/2 \rceil$ parameters. Additionally, the classical post-processing layer contributes $2\times (\lceil n/2 \rceil+1)$ classical parameters. It's worth noting that the number of qubits, $n$, depends on the type of source model chosen, as different source models extract varying lengths of features.\par

\section{Experiments and results}
\subsection{The experimental design}
To thoroughly demonstrate the universality and advantages of the proposed AE-CQTL framework, two widely used toolkits, namely Qiskit \cite{contributors2023qiskit} and PyTorch \cite{paszke2019pytorch}, were employed to conduct image binary classification experiments on the TLQNN and TLQCNN models introduced in Section 3. Furthermore, to comprehensively assess the model performance in a rigorous manner, comparative experiments were conducted with a benchmark classical CCTL model.\par
\begin{figure}[t]
  \centering
  \includegraphics[scale=0.33]{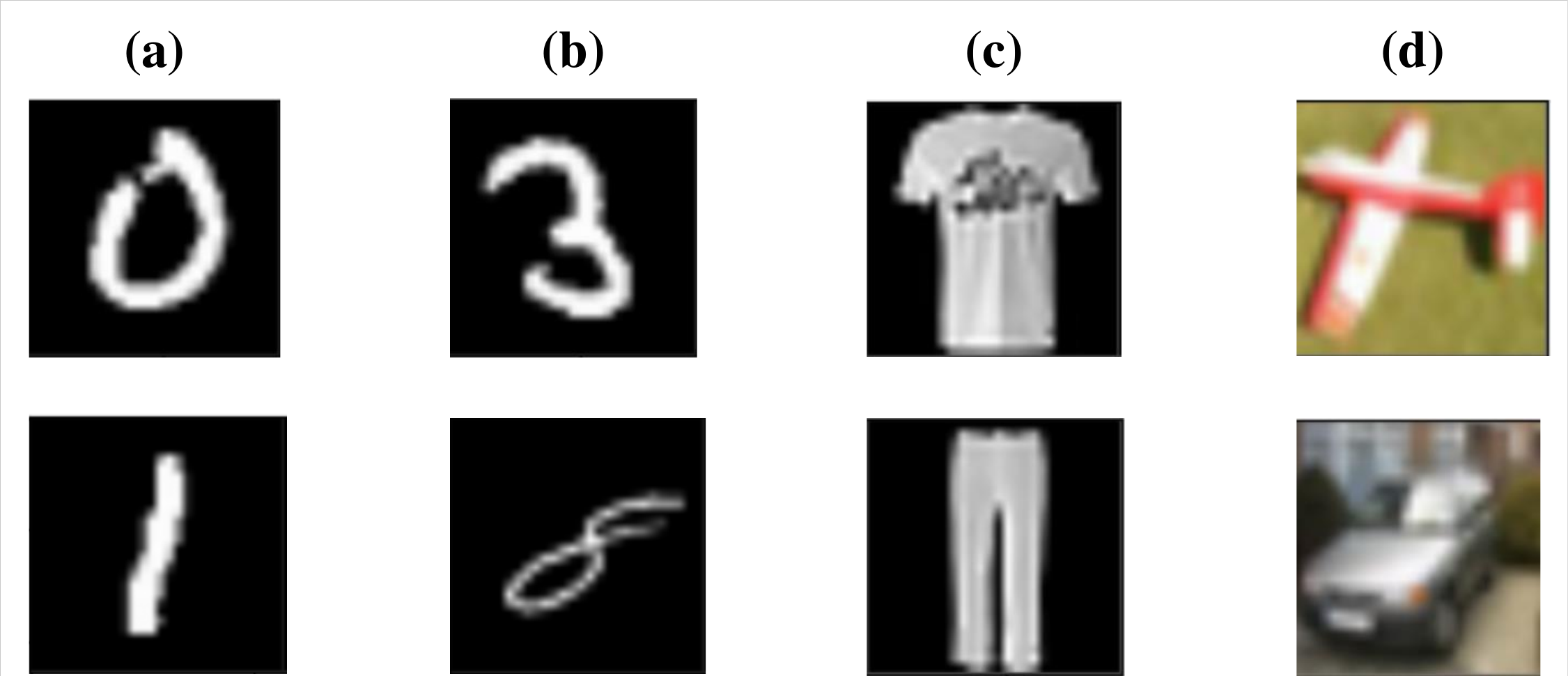}
  \caption{(a) Example of MNIST dataset labeled 0 and 1, corresponding to handwritten digits '0' and '1'. (b) Example of MNIST dataset labeled 3 and 8, corresponding to handwritten digits '3' and '8'. (c) Example labeled 0 and 1 in the Fashion-MNIST dataset, corresponding to T-shirt and trousers. (d) Example of CIFAR10 dataset labeled 0 and 1, corresponding to aircraft and automobiles.}
  \label{fig:8}
\end{figure}

In the selection of our experimental dataset, three benchmark datasets that are widely recognized were chosen: MNIST, Fashion-MNIST, and CIFAR10. For three datasets, binary classification experiments were conducted specifically between category 0 and category 1. Additionally, to increase the difficulty, an additional binary classification task was performed on the MNIST dataset, distinguishing between category 3 and category 8. Images were resized to meet the input dimension requirements of the source model, and normalization was performed to ensure data balance across different dimensions. These preprocessing steps enhance the model's ability to learn and extract meaningful features from the images. Fig.\ref{fig:8} illustrates examples of images from the datasets used.\par
The ResNet18, DenseNet121 and ResNet50 were selected as source models for our experimental setup. In the realm of quantum transfer learning, a pivotal aspect lies in the length of the features extracted by the source model serving as a feature extractor. For instance, ResNet18 extracts features of a length of 512, compatible with 9-qubits PQCs. Analogously, DenseNet121 and ResNet50 extract features with lengths of 1012 and 2048, respectively, corresponding to 10-qubits and 11-qubits PQCs. Notably, all source models employed in our experiments were sourced from the torchvision model library, a renowned repository provided by PyTorch. The incorporation of these three source models serves dual objectives. Firstly, it aims to validate the universality of AE-CQTL models across a diverse range of source models. Secondly, it enables an examination of the trainability of AE-CQTL models with varying qubit counts, specifically ranging from 9 to 11 qubits.\par
\begin{figure}[t]
  \centering
  \includegraphics[scale=0.4]{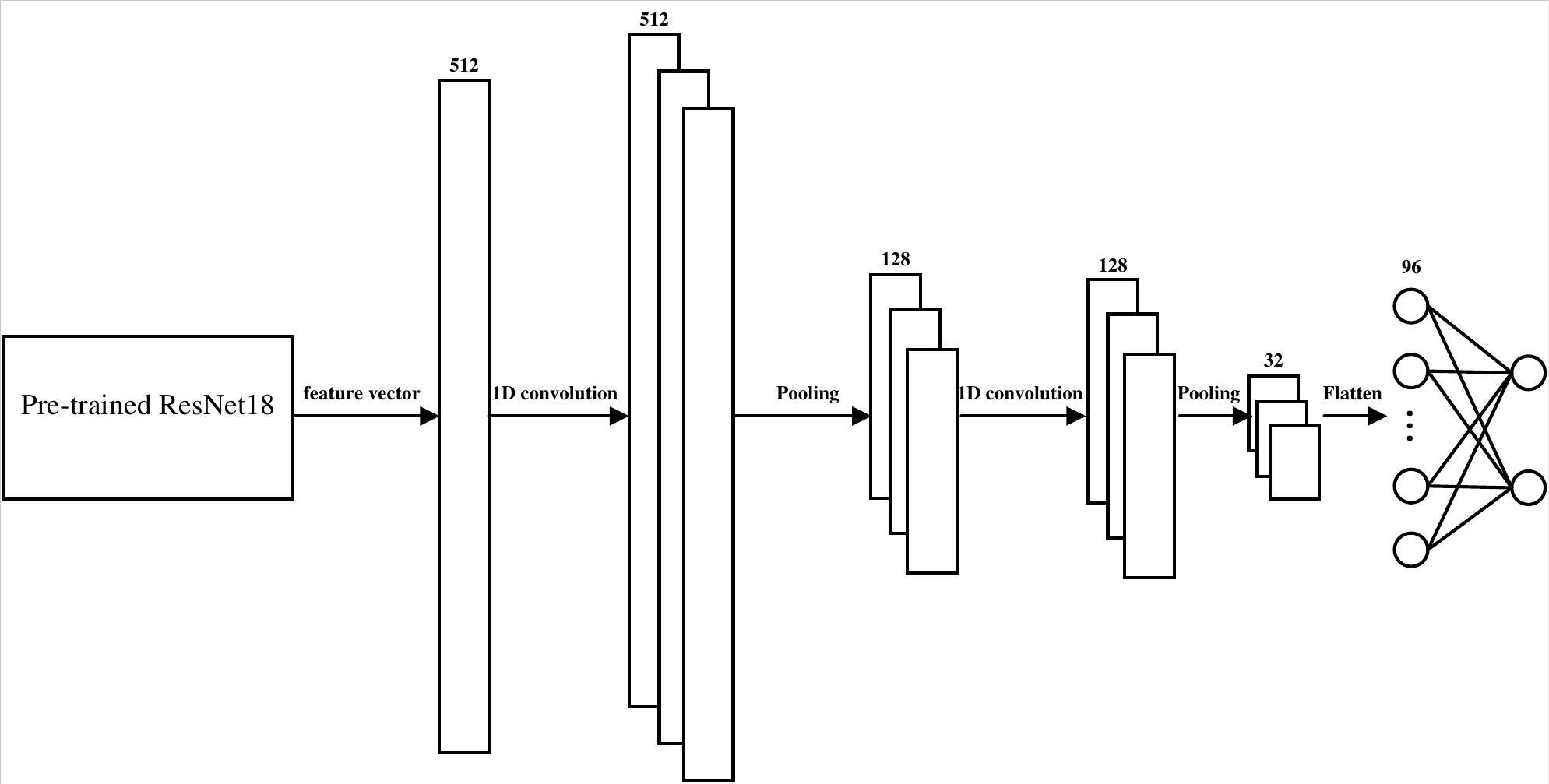}
  \caption{The schematic diagram of the CCTL model based on the ResNet18 source model.}
  \label{fig:9}
\end{figure}

To get a balance between model performance and training time, the number of ansatz layer within the PQCs layer of the TLQNN model was set to 4. Regarding the TLQCNN model, the quantum convolutional and pooling layers were constrained to a single layer each, whereas the quantum fully-connected layer comprised six layers. To evaluate the model performance comprehensively, the proposed models were compared with a benchmark classical CCTL model. When using the same source model as the AE-CQTL models, careful adjustments of CCTL model was made to ensure a comparable number of parameters among all three models. This design allowed us to assess the advantages of the AE-CQTL model over classical models more accurately. Fig.\ref{fig:9} presents a schematic diagram of the CCTL model based on the ResNet18 source model. Similar network structures have been previously employed in other studies of quantum advantage \cite{hur2022quantum,kim2023classical}.\par
For the CCTL model depicted in Fig.\ref{fig:9}, the pre-trained ResNet18 serves as a feature extractor, initially extracting a feature vector of length 512 from the data. Subsequently, the feature vector undergoes two rounds of one-dimensional convolution and pooling operations, reducing its length to 96. Finally, the processed feature vector is fed into a fully-connected layer $\mathcal{L}_{96\rightarrow 2}$ to complete the binary classification task. The number of parameters in the model shown in Fig.\ref{fig:9} is 206, while the corresponding TLQNN and TLQCNN models have 155 and 179 parameters, respectively. The number of parameters for all experimental models is presented in Table 2.\par
\begin{table}[h!]
  \centering
  \caption{The number of parameters for all experimental models}
  \label{tab:parameters}
  \begin{tabular}{lccc} 
      \toprule
      & ResNet18 & DenseNet121 & ResNet50 \\ 
      \midrule
      TLQNN  & 155 & 172 & 189 \\ 
      TLQCNN & 179 & 197 & 221 \\ 
      CCTL  & 206 & 214 & 236 \\ 
      \bottomrule
  \end{tabular}
\end{table}
In each binary classification experiment, 256 samples were selected from each category for training data and 128 samples for testing data to ensure a consistent number of data points for each class. The mini-batch gradient descent algorithm was chosen for parameter optimization, with the cross-entropy loss function employed as the loss metric. Specifically, in each iteration, a batch of 4 samples was randomly selected from the training set to introduce more randomness and help the model escape local optimal. The average gradient for this small batch of samples was then computed and used to update the model parameters. To further improve model performance, we utilized the classical Adam optimizer to optimize both the quantum parameters and the classical parameters in our models. The Adam optimizer is widely employed in deep learning for its simplicity and efficiency. In our experiments, we set the initial learning rate for the Adam optimizer to 0.01. From the first epoch onwards, the learning rate was decayed every 10 epochs with a decay factor of 0.1. All models were trained for a total of 20 epochs, and the experiments were independently repeated 5 times. The hyperparameters and training conditions for the CCTL model remained consistent with those of the AE-CQTL models.\par
\subsection{The experimental results}
In this section, in order to show the excellent performance of the proposed model more comprehensively, we present and compare the experimental results from the aspects of accuracy, stability, convergence, generalization capability, etc.\par
\subsubsection{Accuracy and stability}
\begin{figure}[t]
  \centering
  \includegraphics[scale=0.5]{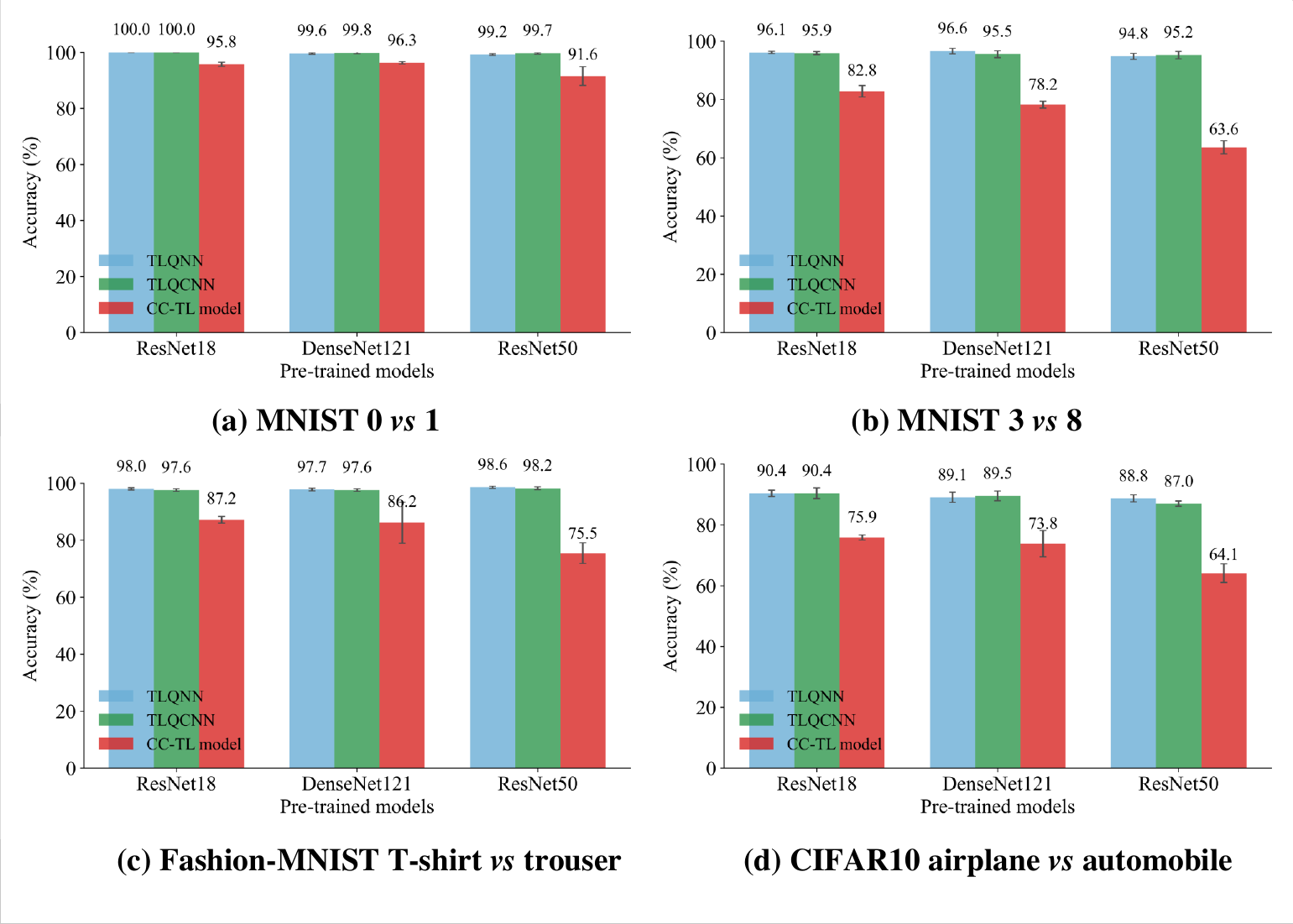}
  \caption{The average accuracy and stability of each model under different datasets. The horizontal axis represents the source model category and the vertical axis represents the model accuracy. The blue bar and the green bar respectively represent the TLQNN model and TLQCNN model under the proposed AE-CQTL framework, and the red bar represents the classical CCTL model.}
  \label{fig:10}
\end{figure}

Fig.\ref{fig:10} exhibits bar charts depicting the mean accuracy of the models across various datasets. Each subfigure corresponds to a distinct dataset, with the horizontal axis enumerating the source model categories in sequential order: ResNet18, DenseNet121, and ResNet50. The vertical axis denotes the accuracy of the models, expressed in percentages. The bars in the figure correspond to different models, specifically, the blue bars signify the TLQNN model, the green bars represent the TLQCNN model, and the red bars indicate the classical CCTL model. The height of each bar indicates the average accuracy of the model across five independent experiments, and the numerical value (rounded to one decimal place) of the average accuracy is displayed above each bar. The error bars represent the standard deviation of the average accuracy, which is a metric used to assess the stability of model under the parameter random initialization strategy. A smaller standard deviation indicates more stable model performance.\par
From Fig.\ref{fig:10}, it is evident that both the TLQNN and TLQCNN models demonstrate excellent performance across various datasets. For instance, in the binary classification experiment of '0' and '1' from the MNIST dataset (Fig.\ref{fig:10}(a)), both models achieve over 99\% accuracy regardless of the source model used. Notably, TLQNN and TLQCNN based on the ResNet18 source model achieve a perfect accuracy of 100\%, and the extremely low standard deviation indicates excellent stability. In contrast, the CCTL model performs slightly worse in terms of both accuracy and stability on this dataset. Its highest accuracy, achieved using the DenseNet source model, is 96.3\%, while the average accuracy obtained using the ResNet50 source model is only 91.6\%. The significant performance variations across different source models suggest that the CCTL model lacks the ability to handle higher-dimensional feature data effectively.\par
When moving to more challenging datasets such as Fashion-MNIST (Fig.
\ref{fig:10}(c)), the AE-CQTL models maintain a high level of accuracy, stabilizing around 98\% across various source models. In contrast, the best performance of the classical model on this dataset is only 87.2\%, with a larger standard deviation. Notably, its performance even dips to 75.5\% when using the ResNet50 source model. Compared to the '0' and '1' classification in MNIST, although the data in Fashion-MNIST is more difficult, the significant decline in performance of the classical model further demonstrates the superiority of AE-CQTL models.\par
In the binary classification experiment of '3' and '8' from the MNIST dataset (Fig.\ref{fig:10}(b)), the increased similarity between the digits poses a greater challenge for classification task. Despite this, TLQNN and TLQCNN maintain accuracy levels around 96\%. In contrast, the performance of the CCTL model experiences a significant decline, with the highest accuracy reaching only 82.8\%. This further validates the superior performance of TLQNN and TLQCNN when dealing with data with high similarity.
In the more complex and challenging experiments on CIFAR10 (Fig.\ref{fig:10}(d)), our two models continue to perform well, achieving an average accuracy of around 90\%. Specifically, the models based on ResNet18 exhibit the best performance, with accuracy hovering around 90.4\%. In contrast, the classical model in the same group only achieves an accuracy of 75.9\%. Therefore, compared to the classical model, TLQNN and TLQCNN demonstrate higher accuracy and stability even on colored datasets.
\subsubsection{Convergence}\par
\begin{figure}[t]
  \centering
  \includegraphics[scale=0.17]{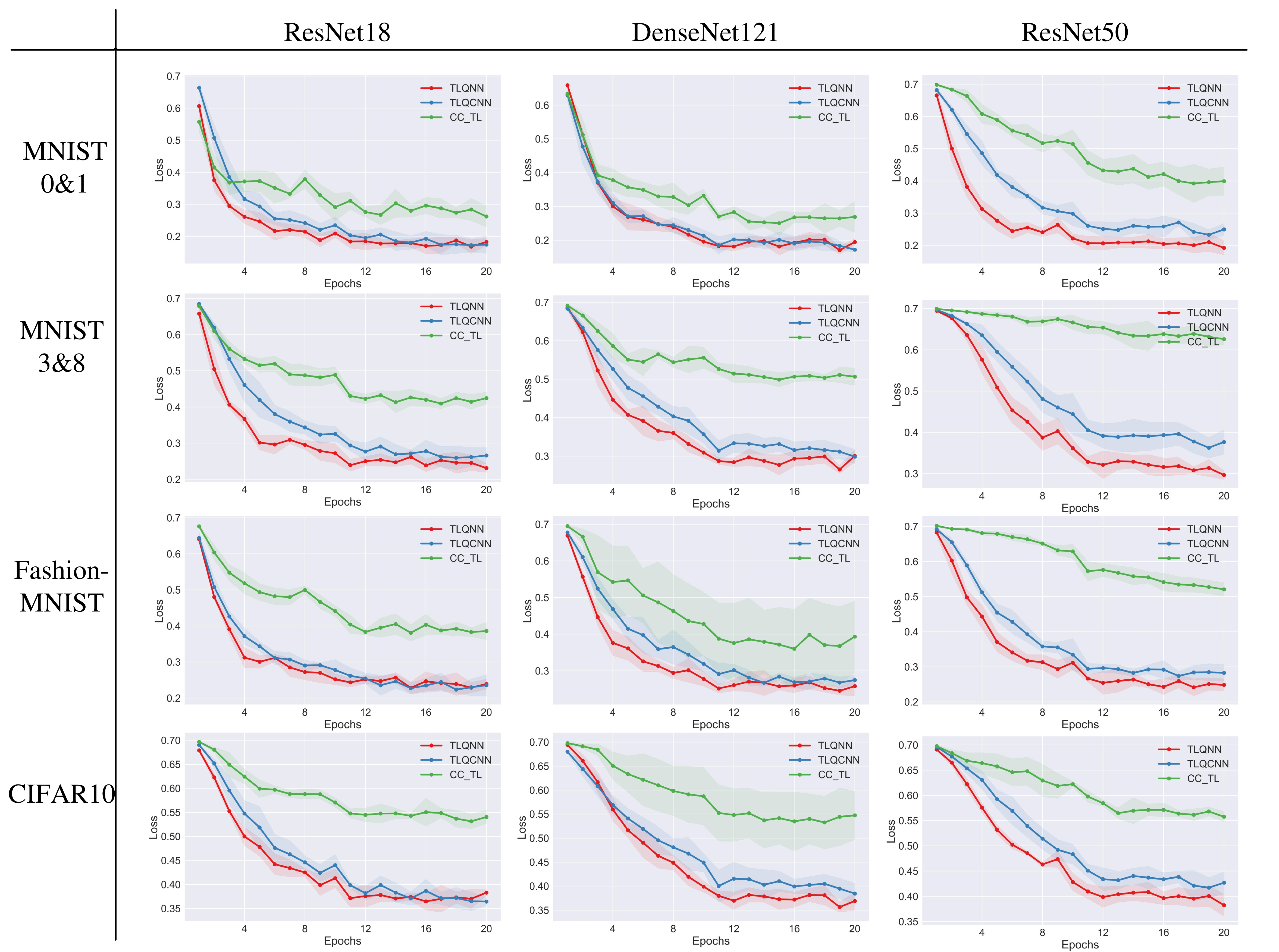}
  \caption{Loss curves of each model. Each subfigure corresponds to a different source model and dataset. The red curve corresponds to the TLQNN model, the blue curve corresponds to the TLQCNN model, and the green curve corresponds to the CCTL model.}
  \label{fig:11}
\end{figure}

To further demonstrate the quantum advantages of our proposed AE-CQTL models, Fig.\ref{fig:11} presents the training loss curves of the models for binary classification experiments across different source models and datasets. It is evident from Fig.\ref{fig:11} that both TLQNN (red curve) and TLQCNN (blue curve) exhibit faster convergence speed and lower converged loss values compared to the classical model (green curve). Notably, the advantages of our proposed models are particularly significant when using ResNet50 as the source model. The phenomenon observed can be attributed to the inherent nature of ResNet50 as a feature extractor, generating features that are significantly longer than those of the other two source models. Specifically, the feature length of ResNet50 is four times longer than that of ResNet18 and twice as long as DenseNet121. Consequently, this imposes a heightened requirement on the model's capability to effectively extract and process these elaborate features. Despite having comparable numbers of parameters, quantum models, owing to their inherent quantum advantages, demonstrate a superior capacity to acquire more pertinent information within an identical training period compared to classical model. This explanation concurs with the findings presented in Fig.\ref{fig:10}, further corroborating the superiority of quantum models in handling intricate datasets.\par
Additionally, the narrower shaded areas in Fig.\ref{fig:11} for the AE-CQTL models reflect their stable performance across multiple repeated experiments, while the classical model exhibits significant fluctuations, as evidenced by the loss curves for the Fashion-MNIST and CIFAR10 datasets using the DenseNet121 source model.\par
\subsubsection{Generalization capability}
\begin{figure}[t]
  \centering
  \includegraphics[scale=0.4]{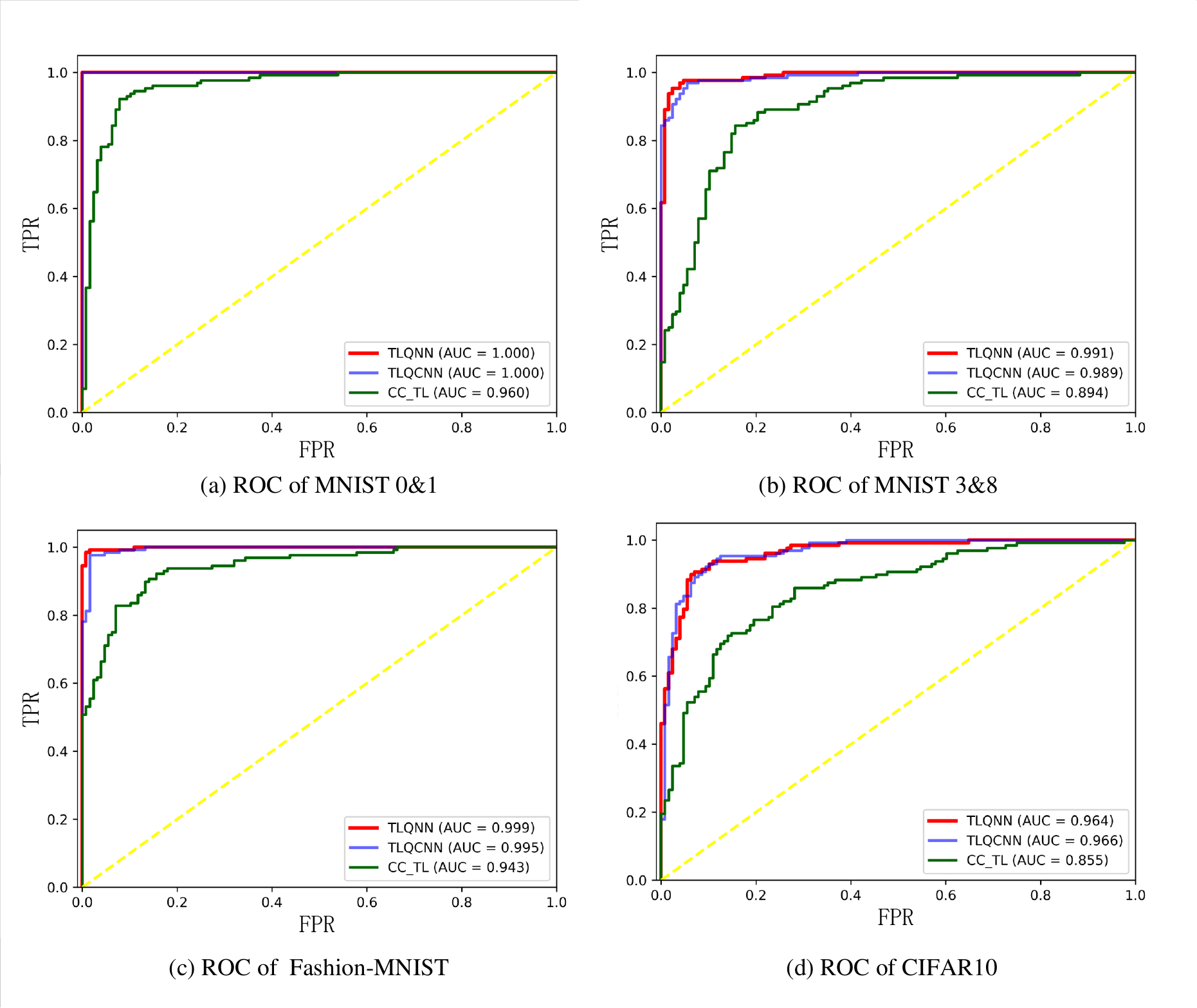}
  \caption{ROC curves for each model. Each subfigure corresponds to a different dataset. The red curve corresponds to the TLQNN model, the blue curve corresponds to the TLQCNN model, the green curve corresponds to the CCTL model, and the yellow dotted line represents the random guess model.}
  \label{fig:12}
\end{figure}

Finally, to compare the generalization capability of TLQNN, TLQCNN, and the classical model across various datasets, ROC curves representative of experiments using ResNet18 as the source model are presented in Fig.\ref{fig:12}. ROC curves are commonly used to assess the generalization capability of ML models. If the ROC curve of one model completely envelopes the ROC curve of another model, it can be concluded that the latter model outperforms the former. In complex scenarios, the area under the ROC curve (AUC) can also be compared, with a larger AUC indicating stronger generalization capability.\par
In Fig.\ref{fig:12}, the vertical axis represents the True Positive Rate (TPR), and the horizontal axis represents the False Positive Rate (FPR). Subfigure (a) corresponds to the binary classification task of handwritten digits '0' and '1' from the MNIST dataset, subfigure (b) corresponds to the binary classification task of handwritten digits '3' and '8' from MNIST, subfigure (c) corresponds to the Fashion-MNIST dataset, and subfigure (d) corresponds to the CIFAR10 dataset. In each subfigure, the red curve represents the TLQNN model, the blue curve represents the TLQCNN model, the green curve represents the classical CCTL model, and the yellow dashed diagonal line represents a random guess model.\par
It can be observed from Fig.\ref{fig:12} that TLQNN and TLQCNN exhibit similar ROC curves across the four datasets, with AUC values indicating that their generalization capabilities are very close. Furthermore, the ROC curves of TLQNN and TLQCNN completely envelope the ROC curve of the classical CCTL model, and the AUC values of the former two models are also higher. This demonstrates that the two AE-CQTL hybrid models proposed in this paper possess higher generalization performance compared to the classical model.\par
\subsection{The experimental summary}
Finally, the experimental results based on the ResNet18 source model were token as the representative, and the performance indicators of each model (accuracy, standard deviation, convergence loss value, AUC) were summarized in Table 3.\par
\begin{table}[h!]
  \centering
  \caption{The performance indicators for all experimental models}
  \label{tab:performance}
  \begin{tabular}{l l c >{\centering\arraybackslash}p{3cm} c}
      \toprule
      & & \textbf{Accuracy (\%)} & \textbf{Convergence loss value} & \textbf{AUC} \\ 
      \midrule
      MNIST (0 vs 1) & TLQNN  & \textbf{100.0 $\pm$ 0.00} & 0.182 & \textbf{1.000} \\ 
                     & TLQCNN & \textbf{100.0 $\pm$ 0.00} & \textbf{0.174} & \textbf{1.000} \\ 
                     & CCTL  & 95.8 $\pm$ 0.75 & 0.262 & 0.960 \\ 
      \midrule
      MNIST (3 vs 8) & TLQNN  & \textbf{96.1 $\pm$ 0.40} & \textbf{0.231} & \textbf{0.991} \\ 
                     & TLQCNN & 95.9 $\pm$ 0.54 & 0.266 & 0.989 \\ 
                     & CCTL  & 82.8 $\pm$ 1.97 & 0.424 & 0.894 \\ 
      \midrule
      Fashion-MNIST & TLQNN  & \textbf{98.0 $\pm$ 0.38} & 0.239 & \textbf{0.999} \\ 
                    & TLQCNN & 97.6 $\pm$ 0.48 & \textbf{0.236} & 0.995 \\ 
                    & CCTL  & 87.2 $\pm$ 1.18 & 0.386 & 0.943 \\ 
      \midrule
      CIFAR10 & TLQNN  & \textbf{90.4 $\pm$ 1.00} & 0.383 & 0.964 \\ 
              & TLQCNN & 90.4 $\pm$ 1.71 & \textbf{0.365} & \textbf{0.966} \\ 
              & CCTL  & 75.9 $\pm$ 0.85 & 0.540 & 0.855 \\ 
      \bottomrule
  \end{tabular}
\end{table}
In summary, the proposed TLQNN and TLQCNN models have demonstrated superior and stable performance compared to the classical classifier across multiple datasets. This includes higher accuracy, lower convergence loss value, and stronger generalization capabilities. Notably, the advantages of these two models are even more evident when dealing with datasets of higher similarity and complexity. This suggests that the proposed AE-CQTL framework is highly feasible and versatile, with the potential for further integration with advanced classical pre-trained deep neural network models. On the other hand, it can be observed that even with a random parameter initialization strategy, TLQNN and TLQCNN models exhibit better trainability and stability in most cases. This is attributed to the rationality of the model structure and the effectiveness of the learning algorithm. Overall, the results indicate that the integration of quantum computing with classical transfer learning offers promising avenues for enhancing the performance of machine learning tasks.\par
\section{Conclusion}
In this paper, an amplitude-encoding-based CQTL framework and the complete learning algorithm were presented, compared to existing frameworks, it employs multi-layer ansatz to multiply the parameters of quantum circuits, providing a basis for deeper classical quantum neural network models. Additionally, two models were designed based on the AE-CQTL framework, namely TLQNN and TLQCNN, specifically for classical image recognition tasks. TLQNN structurally resembles a classical feedforward neural network, utilizing repeated ansatz layers for feature extraction and a classical gradient descent optimizer for parameter optimization. While TLQCNN utilizes quantum convolution and pooling operations, it only needs to perform local Pauli measurements on the retained quantum subsystems after pooling, which effectively alleviates the barren plateau problem during model training. For the two proposed AE-CQTL models, rigorous experimental validations were designed to demonstrate their feasibility and effectiveness. The experimental results confirm that TLQNN and TLQCNN can flexibly integrate with various mature pre-trained deep neural network models, achieving impressive performance on multiple benchmark image datasets. In addition, comparative experiments were conducted between the AE-CQTL models and classical neural network models. Under identical training conditions and comparable parameter counts, the proposed AE-CQTL models significantly surpassed the classical neural network models in terms of accuracy, convergence speed, stability, and generalization capabilities, offering a potential quantum advantage. This work advances the understanding of hybrid quantum-classical systems' scalability thresholds, while simultaneously establishing methodological foundations for engineering quantum algorithms with application-specific efficiency.\par
Indeed, there remain several intriguing questions for future exploration. Firstly, exploring the quantum advantage of AE-CQTL hybrid models on noisy real NISQ devices is also a highly significant research direction. Moreover, the potential reasons for the TLQNN and TLQCNN models show their respective advantages can also be further studied. 
Furthermore, the emergence of large language models naturally raises the question of whether these advanced NLP models can be utilized as source models to construct new AE-CQTL models for various NLP tasks. \par

\section{Acknowledgements}
This work is supported by the National Natural Science Foundation of China (Grant Nos. 61871120 and 62071240), Natural Science Foundation of Jiangsu Province, China (Grant Nos. BK20191259 and BK20220804), Innovation Program for Quantum Science and Technology (Grant No.2021ZD030-2901) and the Jiangsu Funding Program for Excellent Postdoctoral Talent (Grant No. 2022ZB107).\par






\end{document}